\documentclass[12pt]{article}

\usepackage{amssymb,bbm}
\usepackage{amsfonts}
\usepackage{latexsym}
\usepackage{amsmath}
\usepackage{amsthm}
\usepackage{geometry}

\def\lddots{\mathinner{\mkern1mu\raise1pt\hbox{.}\mkern2mu
\raise4pt\hbox{.}\mkern2mu\raise7pt\vbox{\kern7pt\hbox{.}}\mkern1mu}}
\makeatletter
\def\numberbysection{\@addtoreset{equation}{section}
 \def\theequation{\thesection.\arabic{equation}}}
\makeatother

\numberbysection

\newcommand{\be}{\begin{eqnarray}}
\newcommand{\ee}{\end{eqnarray}}

\def\ds{\displaystyle}

\def\bb{\mathbbm}
\def\C{\bb C}

\def\I{\mathbb{I}}

\def\d{\delta}

\title{\bf Reflection $K$-matrices related to Temperley-Lieb $R$-matrices}
\author{ \textsf{Jean Avan$^1$}
\textsf{, Petr Kulish$^2$}
\textsf{ and Genevi\`eve Rollet$^3$}
\\\\
\textit{$^{1,}$ $^3$Laboratoire de Physique Th\'eorique et Mod\'elisation (CNRS UMR 8089),} \\
\textit{Universit\'e de Cergy-Pontoise, F-95302 Cergy-Pontoise, France} \\
\\
\textit{$^2$
St. Petersburg Department of Steklov Mathematical Institute} \\
\textit{Fontanka 27, 191023, St. Petersburg, Russia} }
\date{}

\footnotetext[1]{\tt avan@u-cergy.fr}
\footnotetext[2]{\tt kulish@pdmi.ras.ru}
\footnotetext[3]{\tt rollet@u-cergy.fr}

\begin{document}

\maketitle
\thispagestyle{empty}
\abstract{The general solutions of the reflection equation associated with Temperley-Lieb $R$-matrices are constructed.
Their parametrization is defined and the Hamiltonians of corresponding integrable spin systems are given.}

\bigskip


\clearpage
\newpage

%
%
\section{Introduction}

The Temperley-Lieb algebra $TL_N(q)$~\cite{TL,Ba}, hereafter denoted TL-algebra, plays a central role in the construction and derivation
of quantum integrable models of great interest in statistical mechanics and solid state physics (see e.g. \cite{Ba,PM}).
In particular it is well known that special representations of the TL algebra
give rise to constant solutions $R$ of the Yang-Baxter equation.
From such $R$-matrices one then constructs closed integrable quantum spin chains~\cite{PPK1} on the space of states, 
tensor product of local state spaces $\C^n$,
$\ds {\cal H}=\underset {1}{\overset {N}{\otimes}}\,\C^n$ for any integer $n$. These spin chains are similar to the spin $1/2$ $XXZ$-model.

In order to formulate the generalization of this construction to TL-related {\it open} spin chains one is lead to consider
scalar (i.e. non-operatorial) solutions to the related reflection equation \cite{ESk}. The complete resolution, and
classification of solutions to such equations, are therefore  key issues in the definition of new quantum
integrable models with a symmetry algebra related to TL algebra. Relevance of such quantum systems is manifold and the
associated algebraic structures present several interesting features:
indeed their corresponding quantum algebra ${\cal U}_q(n)$ is different from $U_q(sl(2))$ for $n > 2$,
while the integrals of motion are elements of the algebra $TL_N(q)$~\cite{KMN}.

To construct these spin chains a (reducible) representation of $TL_N(q)$ on  $\ds {\cal H}$  will be used.
Let us be more specific: The TL $R$-matrix considered throughout this paper
is parametrized by an invertible $n\times n$ matrix $b$ while the
parameter $q$ of the corresponding ``$XXZ$-type'' TL spin model is given by:
\be
\label{q}
tr(b^{-1} b^t)=-(q+\frac{1}{q}).
\ee
It was pointed out \cite{PPK1} that the $n\times n$ matrices $K$ solving the reflection
equation for this TL $R$-matrix actually satisfy a quadratic equation:
\be
\label{Kquad}
q\,K^2+\,c_1\,K+(q+\frac{1}{q})^{-1}(c_1^2+q\,c_2) = 0
\ee
with appropriate central elements $c_1$ and $c_2$ depending on $K$ and $b$.

The aim of the paper is to describe a complete parametrization of these $K$-matrices.
In addition, once the constant $R$- and $K$-matrices are known, the Yang-Baxterization
procedure then yields the spectral parameter dependent reflection matrices $K(u)$, which are
cornerstones of the quantum inverse scattering method~\cite{FaTa,Fa,KS,ESk}.

We shall first recall more precisely the context of TL $R$-matrices
construction from braid groups and Hecke algebras, prepare the notation and formulate a derivation of
the quadratic equation (\ref{Kquad}).
The complete classification and full parametrization of constant solutions shall be obtained (Section $3$).
Finally the Yang-Baxterized form for the $K$-matrices will be given. The integrable spin systems
with boundary interactions will be constructed and some comments on their spectral
properties will be given (Section $4$).

\section{Hecke and Temperley-Lieb Algebras} 

Both Hecke algebra $H_N(q)$ and TL  algebra $TL_N(q)$ are quotients
of the group algebra of the braid group  $\mathcal{B}_N$ generated by $(N-1)$ generators
$\check{R}_j$, $j=1,2,\ldots,N-1$ and the relations (see~\cite{CP}):
\begin{eqnarray}
\label{BG}
\check{R}_j\check{R}_k  \check{R}_j  =  \check{R}_k \check{R}_j  \check{R}_k , \ \mbox{for} \ |j-k| =1 \quad \mbox{and}\quad
\check{R}_j\check{R}_k  =  \check{R}_k \check{R}_j , \ \mbox{for} \ |j-k| > 1.
\end{eqnarray}
The Hecke algebra $H_N(q)$ is obtained by adding to these
relations the following constraints  obeyed by each generator $\check{R}_j$ ($q$-deformation of the group 
algebra of the symmetric group):
\be
\label{cheqR}
\left( \check{R}_j - q \right) \left( \check{R_j} + 1/q \right) = 0.
\ee
Equation (\ref{cheqR}) is equivalent to  write $\check{R_j}$ in term of some idempotent $X_j$, namely:
\be
\label{RtoX}
\check{R_j} =  q \mathbb{I} + X_j
\ee
with
\be
\label{Xproj}
X_j^2 =  - \left( q + \frac{1}{q} \right) X_j.
\ee
$\mathbb{I}$ denotes the identity in the Hecke algebra.
The braid group relations (\ref{BG}) read in terms of the idempotents $X_j$ and $X_k$ such that $|j-k|=1$:
\begin{equation}
\label{BGX}
X_jX_kX_j - X_j = X_kX_jX_k - X_k .
\end{equation}
Finally the TL algebra $TL_N(q)$ is obtained as the quotient algebra of the Hecke algebra $H_N(q)$ by
the set of equations requiring that each side of
(\ref{BGX}) be  zero.
To sum up, $TL_N(q)$  is defined by the generators $X_j$, $j=1,2,\ldots, N-1$ and their relations:
\begin{eqnarray}
&&X_j^2 =  - \nu (q) X_j, \nonumber \\
&&X_jX_kX_j = X_j , \quad | j - k | = 1, \nonumber \\
\label{TLX}
&&X_jX_k = X_kX_j , \quad | j - k | > 1
\end{eqnarray}
with $\nu(q)=q+1/q$.

The dimension of the Hecke algebra, $N!$, is  the same as the dimension
of the symmetric group,  whereas the dimension of $TL_N(q)$ is
equal to the Catalan number $C_N=(2N)!/N!(N+1)!$. Implementation of the TL constraint
thus considerably reduces the dimension of the algebra.

In connection with integrable spin systems we will
be interested in representations of $TL_N(q)$ on the tensor product space $\mathcal{H}=\underset {1}{\overset {N}{\otimes}}\,\mathbb{C}^n$.
We will consider in the following, as mentioned in the introduction section,  a particular  representation (reducible)  defined by a single complex invertible $n \times n$ matrix
$b$ which can also be seen as a vector of $\mathbb{C}^n \otimes \mathbb{C}^n$ (with $n^2$
entries: $\{{b}_{cd} \}$)~\cite{PPK1}. We use the notation
$\bar{b}:=b^{-1}$ and view this matrix also as a vector of $\mathbb{C}^n \otimes \mathbb{C}^n$
with entries $\{\bar{b}_{cd} \} $.

The matrix realization on $\mathcal{H}$ of the idempotent generator $X_j$ now reads in terms of $b$:
\begin{equation}
\label{Xbb}
X_j = \underset {j-1}{\underbrace {\I \otimes \ldots\otimes \I}} \otimes
\left (
\sum_{\tiny \begin{array}{c} c,d,c',d\\ \in \{1\ldots n\}\end{array}} b_{cd} \bar{b}_{c'd'} E_{cc'}\otimes E_{dd'}
\right )
\otimes \underset {N-j-1}{\underbrace {\I \otimes \ldots\otimes \I}}
\end{equation}
where we have now denoted by $\I$ the identity matrix in  $End(\C^n)$ and we have used 
the canonical basis of $n \times n$ matrices,
$E_{cc'}$ denoting the $n \times n$ matrix with entries  $(E_{cc'})_{xx'}=\d_{cx}\,\d_{c'x'}$.

One sees here that
$\ds \left ( \sum_{c,d,c',d'=1,\ldots,n} b_{cd} \bar{b}_{c'd'} E_{cc'}\otimes E_{dd'}\right )$ is proportional
to  a rank-$1$ projector
on $\C^n \otimes \C^n$.
Direct computation shows that the set of  relations (\ref{TLX}) are satisfied and they fix the value 
of the parameter $q$  up to a duality $q\rightarrow 1/q$ :
\begin{equation}
\label{trbb}
- \nu (q)=\mathrm{tr} \, \bar{b} b^t = - \left( q + \frac{1}{q} \right).
\end{equation}

The $\check{R_j}$ generators are now also represented in terms of endomorphisms of $\mathcal{H}$. From the particular
form (\ref{Xbb}) these endomorphisms can be consistently denoted as $\check{R}_{j j+1}$. Conditions (\ref{BG}) are
in particular represented as the braided Yang-Baxter equation:
\be
\label{ConstBrYB}
\check{R}_{12}\ \check{R}_{23}\ \check{R}_{12} =  \check{R}_{23}\ \check{R}_{12}\ \check{R}_{23}.
\ee

The $R$-matrix is then defined from this representation of the braid group generators
by $R_{j j+1}={\cal P}_{j j+1} \check{R}_{j j+1}$, with ${\cal P}( v \otimes v' ) = v'\otimes v $ 
for any couple of vectors of $\C^n$. The indexation $j j+1$ of ${\cal P}$ is self-explanatory.
The notation $R_{j j+1}$ is then straightforwardly extended to define general endomorphisms $R_{ij}$ of 
$\mathcal{H}=\otimes_{j=1}^N \ (\mathbb{C}^n)_j$ labeled by any  pair 
of ``site indices'' $(i,j)$, using the time-honored notation \cite{FaTa} for such
elements of $End({\cal H})$ with indices labelling the spaces.

Equation (\ref{ConstBrYB}) then immediately becomes the Yang-Baxter equation for $R$:
\be
\label{ConstYB}
R_{12}\ R_{13}\ R_{23} = R_{23} \ R_{13} \ R_{12}.
\ee

Let us finally formulate the Yang-Baxterization procedure, defined in~\cite{LevMar,AD,KM}, of these $R$-matrices. In fact the Yang-Baxterization procedure
is already valid at the stage of abstract Hecke algebra generators: Indeed if one defines
the spectral parameter-dependant $R$-matrix as
\be
\check{R}_{j}(u)= u \check{R}_{j} - \ds{\frac1u} \check{R}_{j}^{-1} = ( u - \ds{\frac1u}) \check{R}_{j} + \frac{\omega(q)}{u}\mathbb{I};
\;\;\;\; \omega(q) = q - \frac{1}{q}
\label{baxtRm}
\ee
one sees that it obeys the cubic equation in braid group form
with multiplicative spectral parameter $u$ (additive spectral parameter is 
of course obtained as $u \equiv e^{\lambda}$):

\be
\check{R}_j(u)\check{R}_k (uw)  \check{R}_j(w)  =  \check{R}_k(w) \check{R}_j(uw)  \check{R}_k(u) , \ \mbox{for} \ |j-k| =1.
\label{baxtH}
\ee

Now once the
generators $\check R$ of the Hecke algebra $H_N(q)$ itself have been represented as $R$-matrices 
acting on some tensor product of two finite-dimensional vector spaces, this procedure will
immediately give rise to solutions of the non-constant Yang-Baxter equation
with multiplicative spectral parameters:
\be
 R_{12}(u) R_{13}(uw) R_{23}(w)= R_{23}(w) R_{13}(uw) R_{12}(u).
\label{YBSP}  
\ee

\section{Classification of the solutions of the constant reflection equation\label{SecClas}}

We present here a complete classification of the solutions to the constant reflection equation (boundary Yang-Baxter equation)
associated to the TL $R$-matrix $R$ (viewed as an endomorphism of $\mathbb{C}^n \otimes \mathbb{C}^n$)
for any value of $n$. The Yang-Baxterization procedure recalled above  will
then allow to obtain general spectral-parameter dependant
reflection matrices $K(u)$ (section $4.1$) which will then be of direct use for the construction of
TL spin chains (section $4.2$).

Let us first recall how one derives the quadratic equation (\ref{Kquad}) satisfied by matrix solutions $K$
of the constant reflection equation:
\be
\label{ReflEq}
R_{12}\ K_{1}\ R_{21}\ K_{2}= K_{2}\ R_{12}\ K_{1}\ R_{21}
\ee
whenever $R$ is a constant solution of the Yang-Baxter equation, associated with the TL representation obtained in the previous section.

Since the $R$-matrix is obtained from the braid group generators  by $R={\cal P} \check{R}$, (\ref{ReflEq}) can be rewritten:
\be
\label{ReflEqChec}
\check{R}_{12}\ K_{1}\ \check{R}_{12}\ K_{1}= K_{1}\ \check{R}_{12}\ K_{1}\ \check{R}_{12}.
\ee
Using $\check{R} =  q \mathbb{I} + X$, it yields:
\be
\label{eqKX}
q\ X_{12}\ K_{1}^2 + \ X_{12}\ K_{1}\ X_{12}\ K_{1}= q\ K_{1}^2 \ X_{12} + K_{1}\ X_{12}\ K_{1}\ X_{12}.
\ee
This equation reads in term of the $b$ matrix:
\be
\label{eqKb}
q \ b \otimes ((K^2)^t \,\,\bar{b}) \ + \ tr( K\, b \bar{b}^t) \ b \otimes (K^t\, \,\bar{b})
=
q \ (K^2 \,b) \otimes \bar{b} \ + \ tr(K\, b \bar{b}^t) \ (K \,b) \otimes \bar{b}.
\ee
Since matrices $b$ and $\bar{b}$ are invertible, this is equivalent (after taking the transposition) to:
\be
\label{eqK}
\I \otimes \ (q \ K^2  + \ tr( K\, b \bar{b}^t)\ K )^t
=
(q \ K^2  + \ tr( K\, b \bar{b}^t) \ K ) \otimes \I.
\ee
This establishes that $ q \ K^2  + \ tr( K\, b  \bar{b}^t)\ K $ is proportional to identity.

The value of the coefficient of the identity term is immediately obtained as a
consistency condition for the value of the linear form ($q$-trace) $ tr (\bar{b} -- b^t)$ 
applied to both sides of the equality \ref{eqK}.
One finally gets the normalized quadratic polynomial annihilating $K$ as:
\be
 K^2  + \frac{1}{q} \ tr(K\, b  \bar{b}^t)\ K = k_2 \mathbb{I}
\label{quadr}
\ee
with
\be
k_2 = \frac{1}{q \,tr ( b \bar{b}^t)} (tr(K\, b \bar{b}^t)^2 + q tr ( K^2\, b \bar{b}^t )).
\label{c2}
\ee

It follows that the complete resolution of the reflection equation for these constant TL
$R$-matrices will be realized in two steps:

$1$. Parametrize all matrices $K$ with a minimal polynomial of degree $2$ (or less).

$2$. Fix the value of the coefficient of the linear term to its expression in (\ref{quadr}).

The coefficient of the constant term, as we have seen, is the result of a self-consistent
evaluation of a trace and therefore does not represent a supplementary independant constraint on $K$.
These two steps are thus necessary AND sufficient to obtain all solutions of the reflection equation.

Step $1$ is separated into three obvious subcases:

$1a$: Minimal polynomial of degree $1$. The matrix $K$ is then proportional to the Identity and automatically
solves the reflection equation without further conditions.

$1b$: Minimal polynomial of degree $2$ with two distinct roots. The matrix $K$ is then diagonalizable
with the same two roots as eigenvalues.

$1c$: Minimal polynomial of degree $2$ with a double root. The matrix $K$ is then only trigonalizable (i.e. is written
with Jordanian cells) with a single eigenvalue and an order-$2$ nilpotency on the corresponding eigenspace.

We now consider in detail cases $1b$ and $1c$.

\subsection{Diagonalizable $K$-matrices}

Any diagonalizable $n \times n$ matrix with two distinct eigenvalues denoted $\lambda$ and $\mu$ can be
parametrized as follows:

\be
K = \lambda \mathbb{I} + (\mu - \lambda) P
\label{form1}
\ee
where $P$ is the projector parallel to the eigenspace
$V_{\lambda}$ with eigenvalue $\lambda$, onto the eigenspace $V_{\mu}$ with eigenvalue $\mu$.
One can always choose $\mu$ such that the dimension of $V_{\mu}$,
hereafter denoted $m$, is lower than (or at most equal to) the dimension of $V_{\lambda}$, hence $m \leq [\frac{n}{2}]$.

The projector $P$ is then constructed from two sets of data encapsulating all the information on $V_{\mu}$
and $V_{\lambda}$ albeit with redundancies:

a: a set of $m$ independent vectors building a basis of $V_{\mu}$, defining in this way an $n \times m$ rectangular
matrix $B$ of maximal rank $m$. The redundancy in this parametrization correspond to the arbitrariness in
the choice of the basis in $V_{\mu}$ , described by the transformation $B \rightarrow Bg$ for any $g$
in $Gl(m)$.

b: a set of $m$ independent vectors building a basis of $\bar{V_{\lambda}}$ defined as the $m$-dimensional vector space
of solutions to the rank $n-m$ homogeneous linear system:

\be
v^t\, C = 0
\label{sysVl}
\ee
where $v$ is the unknown $n$ dimensional vector and $C$ is an $n \times (n-m)$ rectangular matrix defined from any basis of
vectors for $V_{\lambda}$ in the same way as $B$ is defined for $V_{\mu}$. $C$ is of course defined up to
rhs multiplication by $g'$ in $Gl(n-m)$ which does not affect $v$. This second set of $m$ $n$-dimensional vectors allows then
to build a second $n \times m$ rectangular matrix $A$ of maximal rank $m$. $A$ is also defined up to
a rhs multiplication by any $h$ in $Gl(m)$. From (\ref{sysVl}) one sees that
$A^t\, \,C = 0$.

In addition one must impose that the intersection of $V_{\mu}$ and $V_{\lambda}$ is empty, which is equivalent
to asking that no vector of $V_{\mu}$ be a solution of (\ref{sysVl}), or finally to requiring that the square
$m \times m$ matrix $A^t\, \,B$ be invertible.

$P$ is then built as:

\be
P = B\,\, (A^t\, \,B)^{-1}\,\, A^t
\label{proj1}
\ee
as is immediately checked by operating $P$ on $B$ (vectors of $V_{\mu}$), yielding again $B$,
 and $C$ (vectors of $V_{\lambda}$), yielding 0.

We recall that this parametrization is defined up to separate rhs multiplication of $A$ and $B$ by any matrix
of $Gl(m)$. This redundancy shall be presently used to simplify (\ref{proj1}).

Hence, any diagonalizable $K$-matrix with $2$ eigenvalues can be written as:

\be
K = \lambda \mathbb{I} + (\mu - \lambda) B\,\, (A^t\, \,B)^{-1}\,\,A^t.
\label{kmat1}
\ee

The number of relevant parameters is thus $2$ (eigenvalues) $+ 2nm$ (matrices $A$ and $B$) $-2m^2$ ($2$ changes of basis
in $Gl(m)$). The redundancy of the parametrization under $A \rightarrow Ah$ and $B \rightarrow Bg$ is manifest
in (\ref{kmat1}).

We now realize Step $2$ by imposing that the value of the coefficient of the linear term
in the minimal polynomial, which is equal to the sum of the two zeroes $\lambda + \mu$, be
identified to its expression in (\ref{quadr}), i.e.:

\be
\lambda + \mu = - \frac{1}{q} tr( K\, b \bar{b}^t)
\label{const1}
\ee
that is:
\be
\lambda ( 1 + \frac{1}{q} tr (b \bar{b}^t) - \frac{1}{q} tr (\bar{b}^t\,\, B\,\, (A^t\, \,B)^{-1}\,\, A^t\,\,b)) +
\mu (1 + \frac{1}{q} tr (\bar{b}^t\,\, B\,\, (A^t\, \,B)^{-1}\,\, A^t\,\,b) = 0.
\label{const2}
\ee
This fixes univocally the ratio $\ds \frac{\lambda}{\mu}$ unless both 
coefficients in (\ref{const2}) vanish. This vanishing in turn implies that 
$tr(b \bar{b}^t) = -2q$ hence from the TL trace condition (\ref{trbb}) $q=\pm 1$. In this case $\check{R}$ is triangular
and indeed no condition may relate the eigenvalues (see e.g. \cite{KMN2}).

We can now use the arbitrariness in the choice of $A$ and $B$ to impose that the $m\times m$ invertible matrix $(A^t\,\,B)$ be set to
$\mathbb{I}$. This leaves
a set of $m^2$ non-relevant parameters. $K$ is then given as:

\be
K = \lambda \{ \mathbb{I} + \{ \frac{-q + 1/q}{q + tr (\bar{b}^t\,\, B\, A^t\,\, b)} \}\, B\, A^t  \}; \;\;\;\; (A^t\,\, B) = \mathbb{I}.
\label{Kdiag}
\ee

The overall number of relevant parameters in $K$ is now $2(n-m)m +1$ (except in the triangular case where one extra
parameter occurs as just seen). Using a spin notation which is useful in the context of
spin chain construction using TL algebra, one equivalently rewrites $n=2s+1$ leading to $2(2s+1 -m)m +1$ parameters.
The irrelevant parameters are the components of the diagonal global $Gl(m)$ gauge transformation $A \rightarrow A\, (g^{-1})^t \;\; ; \;\;
B \rightarrow Bg$.

This enables us to identify the solutions recently proposed in \cite{LS} (at least the infinite-spectral parameter limit
thereof, which solve the constant reflection equation). They are precisely the diagonalizable solutions corresponding to the choice
$\ds m = [\frac{2s+1}{2}]$. For instance when $s = 3/2$ ($n=4$) one obtains a $9$ parameter solution which can be shown
to have two eigenvalues of multiplicity $2$. Explicit formulation of the eigenvectors is also available but the particular
form of the parametrization used in \cite{LS} yields cumbersome formulae which we shall not give here.

\subsection{Non-diagonalizable $K$-matrices}

In this case $K$ is automatically of the form $\lambda \mathbb{I} + N$ where $N$ is a nilpotent $n \times n$ matrix: $N^2 = 0$. A similar
parametrization for $N$ as in the diagonalizable case exists. Set the dimension of the kernel of $N$
to be $n-m$ with of course $m \leq  [n/2]$ since $Im N \subset Ker N$.  This time the $m$-dimensional image of the cokernel of $N$ yields a
rectangular matrix $B$ up to rhs multiplication by $g$ in $Gl(m)$. The $(n-m)$-dimensional kernel of $N$, as in (\ref{sysVl})
can again be characterized by another $n \times m$ rectangular matrix $A$. However this time one must impose a complete inclusion condition
of the image vectors defining $B$ in the kernel, in other words $A^t\,\, B = 0$. $N$ is then immediately obtained as $N = B\,A^t$. Because
of the condition $A^t\,\, B = 0$ the scale of $N$ is not fixed; this scale fixing is here obtained by the implementation of Step $2$
to impose:

\be
\lambda (q - q^{-1}) = tr (\bar{b}^t\, B\, A^t\,\, b).
\label{step2}
\ee

The irrelevant parameters are again the components of the diagonal global $Gl(m)$ gauge transformation $A \rightarrow A\, (g^{-1})^t \;\; ; \;\;
B \rightarrow Bg$. The number of relevant parameters is thus $1$ (eigenvalue) $+ 2nm$ (matrices $A$ and $B$) $-m^2$ (changes of basis
in $Gl(m)$) $-m^2$ (inclusion relation $A^t\,\, B = 0$) $-1$ (trace relation) $= 2nm - 2m^2$.

\subsection{Complete parametrization}

Both situations can now be summarized into a single representation:

{\bf Proposition}

Any solution to the constant reflection equation (\ref{ReflEq}) related with $TL$-algebra, takes the form:

\be
K = \lambda \mathbb{I} + B\, A^t
\label{genparam}
\ee
where $A$ and $B$ are rectangular $n \times m$ matrices of rank $m$, 
$m\leq[\ds \frac{n}{2}]$  defined up to a diagonal $Gl(m$) gauge 
transformation $g$:

\be
A \rightarrow A\, (g^{-1})^t \;\; ; \;\; B\rightarrow Bg
\label{gauge}
\ee
and submitted to the condition:

\be
A^t\,\, B = (\mu - \lambda) \mathbb{I}
\label{condd}
\ee
and
\be
-\frac{1}{q}\lambda + q \mu = - tr (\bar{b}^t\,\, B\,A^t\,\, b).
\label{norm2}
\ee

If $\mu = \lambda$ one recovers the non-diagonalizable case

If $\mu \neq \lambda$ one recovers the diagonalizable case.

\section{Spectral parameter dependent  K matrices}

\subsection{Yang-Baxterization}

The Yang-Baxterization of the Hecke $R$-matrices was formulated
in Section $2$, eqn. (\ref{baxtRm}). In order to define the corresponding Yang-Baxterization for the associated
$K$-matrices obeying the reflection equation one is lead to define the extension
of the Hecke algebra to the affine Hecke algebra $\hat H_N(q)$. It has one more generator $K$ with relations:
 \be
  \check{R}_{1}K \check{R}_{1}K = K \check{R}_{1}K \check{R}_{1} \quad  \mbox{and} \quad
 K \check{R}_{j} = \check{R}_{j}K \; \mbox{for all} \; j > 1.
\label{affH}
\ee

As in the Hecke case an extra polynomial constraint imposed on $K$ as $p_n(K) = 0$, will define
a quotient of $\hat H_N(q)$ known as ``cyclotomic Hecke algebra'' \cite{Ari}. There exists then consistent realizations of the Yang-Baxterized
$K$-matrix by Laurent polynomials $K(u)$ in $ u, u^{-1}$, depending on the coefficients of $p_n$ and $K^m$,
$m= 0, 1, ... n-1$ \cite{AD,KM}. They are solutions to the algebraic reflection equation \cite{ESk}:
\be
 \check{R_{1}}(u/w) K(u) \check R_{1}(uw) K(w) =
  K(w) \check R_{1}(uw) K(u) \check R_{1}(u/w)
\label{SPRA}
\ee
where $\check R_{1}(u/w)$ is the Yang-Baxterized $\check{R}$ matrix (\ref{baxtRm}). Once suitable matrix representations of
both $R$ and $K$ are considered, respectively in $ End(\C^n \otimes \C^n)$ and $End(\C^n)$ this becomes the well-known braided 
Sklyanin reflection equation
\be
\check{R_{12}}(u/w) K_1(u) \check R_{12}(uw) K_1(w) =
  K_1(w) \check R_{12}(uw) K_1(u) \check R_{12}(u/w).
\label{SPRA2}
\ee

We are here considering specifically the case of a represented TL-type $R$-matrix built from a rank-$1$ projector.
We have seen in Section $3$ that the constant solution of the reflection equation then satisfies a quadratic constraint.
It follows from general arguments \cite{LevMar,AD,KM} that the corresponding spectral parameter-dependent $K(u)$ is given by
the expression:
\be
 K(u) = u^2 K - \frac{1}{u^{2}} K^{-1} + c 
\label{BaxK}
\ee
with an arbitrary central element $c$. Due to the relation (\ref{BaxK}), after
a suitable normalization of $K$, one gets the regularity property: $ K(u)| (u=1) =  \mathbb{I}$.
This property is important to construct an integrable spin chain Hamiltonian
with nearest-neighbour interaction and a boundary interaction on the
left and right boundary sites described by matrices $K^-(u)$ and $K^+(u)$
respectively \cite{ESk}.

\subsection{Spin chains}

The construction of the spin chain Hamiltonian proceeds now from general principles. Taking the $R$-matrix
$R_{0j}(u) = {\mathcal P}_{0j} \check{R}_{0j}$ as an $L$-operator at each site $j$ with auxiliary space labeled by index $0$, one
constructs the monodromy matrix \cite{FaTa,Fa}:
\be
T(u) = L_{0N}(u) L_{0 N-1}(u) \cdots L_{01}(u)
\label{monod}
\ee
and the two-row monodromy matrix \cite{ESk}:
\be
{\cal T}(u)= T(u) K^{-}_0 (u) T^{-1}(1/u)
\label{rowmonod}
\ee
where $K^{-}_0 (u)$ is a solution of the reflection equation.
The generating function of integrals of motions (including the Hamiltonian) is:
\be
\tau(u) = tr K^{+}_0 (u) T(u) K^{-}_0 (u) T^{-1}(1/u).
\label{tau}
\ee
where in addition $K^{+}_0 (u)$ is a solution of the suitably 
defined dual reflection equation. All solutions thereof can be obtained 
straightforwardly from the set of solutions $K^-$ due to crossing-unitarity 
of $R$-matrix. In terms of the $R$-matrix $\tau(u)$  reads:
\be
\tau (u) = tr K^{+}_0 (u) R_{0N}(u) R_{0 N-1}(u) \cdots 
R_{01}(u) K^{-}_0 (u) R_{10}(u) R_{20}(u) \cdots R_{N0}(u).
\label{tauR}
\ee
With an appropriate normalization one fixes $\tau(1) = tr K^{+}_0 (1)$ (remember that in the multiplicative
spectral parameter representation the critical value for regularity properties is $1$). The spin chain Hamiltonian
becomes then proportional to the local expression:

\be
H = \sum_{k=1}^{N-1} \frac{d}{du}\check{R}_{k k+1}(u=1) + \frac{1}{2} \frac{d}{du} K^{-}_1 (u=1)
+ (tr K^{+}_0 (1))^{-1} tr K^{+}_0 (1) \frac{d}{du}\check{R}_{N 0}(u=1)
\label{hamspin}
\ee
where the contribution of the boundary conditions is explicit.

If one chooses general $n\times n$ $K$-matrices $K^+$ and $K^-$ with the full arbitrariness parametrized
in Section $3$ this contribution makes it difficult to characterize quantitatively properties (such as spectrum
and eigenvectors) of the spin system. At this time we lack a proper framework to deal with this
general case and this is a key issue which must be adressed in the future.

To illustrate what could be done, when the suitable algebraic tools available, 
let us finally concentrate on the simplest particular case which indeed 
can be treated very extensively using general algebraic arguments.

Restricting oneself to the free ends case \cite{MN}:
\be
K^{-}_1 (u) = \mathbb{I} \;\;\;\;\;K^+_1 (u) = M =  b^t\,\,\bar{b}
\label{free}
\ee
where $M$ is the matrix entering into the crossing-unitarity relation for $R$, the spin chain Hamiltonian and
the higher conserved quantities then lose altogether their boundary contributions and become elements of the TL algebra,
for instance:
\be
 H = \sum_{k=1}^{N-1} \frac{d}{du}\check{R}_{k k+1}(u=1) \in TL_N(q)
\label{TLHam}
\ee
This Hamiltonian is now symmetric w.r.t. the quantum algebra ${\mathcal U}_q(n)$ and can be restricted to the irreducible
representation subspaces of $TL_N(q)$ in a decomposition of the phase space:
\be
 \mathcal{H}=\underset {1}{\overset {N}{\otimes}}\, \mathbb{C}^n = \underset {k=0}{\overset {[N/2]}{\oplus}}W_k \otimes \mathbb{C}^{\nu(k)}
\label{decomp}
\ee
where $W_k$ denotes the irrep of $TL_N(q)$ corresponding to the two-row Young diagramme with partition
$\{ (\lambda_1, \lambda_2)| \lambda_1 + \lambda_2 = N, \lambda_2 = k \} $ and $\nu(k)$ is the multiplicity
of this irrep in the decomposition. Hence the spectrum of $H$ consists here of multiplets of subspaces
$\{E^{(j)}_k\}, j= 1, 2, \cdots dim W_k$. associated with the irreps $W_k$, each with multiplicity $\nu(k)$.

\vspace{.75cm}
{\bf Acknowledgements}
\vspace{.5cm}

This work was sponsored by CNRS; Universit\'e de Cergy-Pontoise; and ANR Project DIADEMS (Programme Blanc ANR SIMI$1$ $2010$-BLAN-$0120$-$02$). 
PPK is partially supported by
GDRI-471 "Formation et recherche en physique th\'eorique" and RFBR grant 09-01-00504-a and 09-01-93108.


\begin{thebibliography}{99}


\bibitem{TL} H.N.V. Temperley, E. Lieb; \emph{Relations between the 'Percolation' and 'Colouring' Problem 
and other Graph-Theoretical Problems Associated with Regular Planar Lattices: Some Exact Results 
for the 'Percolation' Problem}, Proc. Roy. Soc. {\bf A 322} (1971), 251.


\bibitem{Ba} R.J. Baxter; \emph{Exactly Solved Models in Statistical Mechanics}, London, Academic Press (1982)

\bibitem{PM} P. Martin; \emph{Potts models and related problems in Statistical Mechanics}, 
World Scientific, Singapore (1991)

\bibitem{PPK1} P.P. Kulish; \emph{On spin systems related to Temperley-Lieb algebra}, 
J. Phys. A (Math.Gen.) {\bf 36} (2003), L489.

\bibitem{ESk} E.K. Sklyanin; \emph{Boundary conditions for integrable quantum systems}, 
J. Phys. A (Math. Gen.) {\bf 21} (1988), 2375.

\bibitem{KMN} P.P. Kulish, N. Manojlovic, Z. Nagy; \emph{Symmetries of spin systems and Birman-Wenzl-Murakami algebra}, 
Journ. Math. Phys. {\bf 49} (2008), 023510.

\bibitem{FaTa} L.D. Faddeev, L.M. Takhtadzyan; \emph{The quantum method for the Inverse Problem and 
the XYZ Heisenberg model}, Usp. Math. Nauk {\bf 34} (1979), 13. 

\bibitem{Fa} L.D. Faddeev; \emph{How the algebraic Bethe Ansatx works for integrable models} 
in \emph{Quantum Symmetries: Proceedings of Les Houches Summer School Session LXIV} 
(1998), 149, ed. by A. Connes, K. Gawedzki and J. Zinn-Justin, North Holland.

\bibitem{KS} P.P. Kulish, E.K. Sklyanin; \emph{Quantum Spectral Transform Methods: Recent Developments} 
in \emph{Integrable Quantum Field Theories}, Lecture Notes in Physics {\bf 151} (1982), 61, 
edited by J. Hietarinta and C. Montonen, Springer.

\bibitem{CP} V. Chari, A.N. Pressley; \emph{A Guide to Quantum Groups}, Cambridge University Press (1995).

\bibitem{LevMar} D. Levy, P. Martin; \emph{Hecke algebra solutions to the reflection equations}, 
J.Phys. A (Math. Gen) {\bf 27} (1994), 521.

\bibitem{AD} A. Doikou; \emph{From affine Hecke algebras to boundary symmetries}, Nucl.Phys. {\bf B725} (2005), 493.

\bibitem{KM}  P.P. Kulish, A. Mudrov;
\emph{Baxterization of solutions to the reflection equation with Hecke $R$-matrix}, 
Lett. Math. Phys. {\bf 75} (2006), 151.



\bibitem{KMN2} P.P. Kulish, N. Manojlovic, Z. Nagy; \emph{Jordanian deformation of the open XXX-spin chain},
Theor. Math. Phys.{\bf 163} No. 2 (2010), 644.

\bibitem{LS} A. Lima-Santos; \emph{On the ${\cal{U}}_{q}[sl(2)]$ Temperley-Lieb reflection matrices},
arXiv:1011.2891v1 [nlin.SI].


\bibitem{Ari} S. Ariki; \emph{Lectures on cyclotomic algebras}, in \emph{Quantum groups and
Lie theory}, London Mathematical Society Lecture Notes Series {\bf 290} (2002), 1, ed. by Andrew Pressley; 
arXiv: math/9908005.

\bibitem{MN}  L. Mezincescu, R.I. Nepomechie; \emph{Integrability of open spin chains with quantum algebra symmetry}, 
Int. J. Mod. Phys. {\bf A6} (1991), 5231; Int. J. Mod. Phys. {\bf A7} (1992), 5657 (Addendum); arXiv: hep-th/9206047. 

\end{thebibliography}
\end{document}